# Maximizing NMR Sensitivity: A Guide to Receiver Gain Adjustment

Josh P. Peters[1*], Frank D. Sönnichsen[2], Jan-Bernd Hövener[1], Andrey N. Pravdivtsev[1*]

1. Section Biomedical Imaging, Molecular Imaging North Competence Center (MOIN CC), Department of Radiology and Neuroradiology, University Medical Center Kiel, Kiel University, Am Botanischen Garten 14, 24114, Kiel, Germany

2. Otto Diels Institute for Organic Chemistry, Kiel University, Otto Hahn Platz 4, 24118, Kiel, Germany

*Corresponding Authors: josh.peters@rad.uni-kiel.de; andrey.pravdivtsev@rad.uni-kiel.de

## Abstract

Novel methods and technology drive the rapid advances of nuclear magnetic resonance (NMR). The primary objective of developing novel hardware is to improve sensitivity and reliability (and possibly to reduce cost). Automation has made NMR much more convenient, but it may lead to trusting the algorithms without regular checks. In this contribution, we analyzed the signal and signal-to-noise ratio (SNR) as a function of the receiver gain (RG) for $^1$H, $^2$H, $^{13}$C, and $^{15}$N nuclei on five spectrometers. On a 1 T benchtop spectrometer (Spinsolve, Magritek), the SNR showed the expected increase as a function of RG. Still, the $^1$H and $^{13}$C signal amplitudes deviated by up to 50% from supposedly RG-independent signal intensities. On 7, 9.4, 11.7, and 14.1 T spectrometers (Avance Neo, Bruker), the signal intensity increases linearly with RG as expected, but surprisingly a drastic drop of SNR is observed for some X-nuclei and fields. For example, while RG = 18 provided a $^{13}$C SNR similar to that at a maximum RG of 101 at 9.4 T, at RG = 20.2 the determined SNR was 32% lower. The SNR figures are strongly system and resonance frequency dependent. Our findings suggest that NMR users should test the specific spectrometer behavior to obtain optimum SNR for their experiments, as automatic RG adjustment does not account for the observed characteristics. In addition, we provide a method to estimate optimal settings for thermally and hyperpolarized samples of a chosen concentration, polarization, and flip angle, which provide a high SNR and avoid ADC-overflow artefacts.

## Introduction

Nuclear magnetic resonance is a universal analytical method[1]. The development of stronger magnets, faster, more precise, less noisy electronics, and more sensitive probes has extended the scope of the technique further[2,3]. Still, the sensitivity of NMR remains limited, so various signal enhancement methods are being developed, including the hyperpolarization of nuclear spins[4–6].

Maximizing the signal-to-noise ratio (SNR) is a nontrivial task. A parameter adjusted for a scan is the receiver gain (RG), which matches the dynamic range of the signal recorder to the strength of the expected and subsequently detected signal. The NMR receiver typically comprises a series of high-performance analog amplifiers, which are activated or deactivated via RG controls. While individual amplifiers generally exhibit similar performance characteristics, accurately predefining the actual gain of a specific amplifier is challenging, and the nominal RG value may not perfectly correlate with the actual gain.[7] However, this can be corrected, for example, on Bruker Neo systems using an RG cortab.

Typically, it is advised to use the automatic adjustment that maximizes RG while avoiding the signal overflow (receiver range threshold, RRT) and, thus, a cut-off or clipping of the free induction decay (FID, **Figure 1**). Clipping leads to strong spectral artifacts, which renders the spectrum useless. However, in the case of hyperpolarization, automated RG adjustment is typically not possible, as the signal is enhanced only transiently and rapidly decays. Therefore, sufficiently low RG and excitation pulses are used to avoid the overflow. However, this approach may result in a loss of SNR if the RG was chosen too low, or results in a clipped FID and failed experiment if the RG was set too high.

Several works studied the linearity of actual RG and its implication to quantitative NMR spectral analysis.[7–9] For optimum sensitivity, a dilute analyte is typically observed with a high RG, and the strong, interfering solvent signal has to be suppressed. On some NMR systems, reliable signal quantification was demonstrated only when the FID amplitude was below 50% of RRT; if the signal was higher, apparent signal amplitude and shape distortions caused by signal compression were observed.[10] Note, that in this work a different hardware generation was used, and the RRT of more modern hardware might be higher.

For quite a while, one of the leading manufacturers of NMR spectrometers, Bruker, has introduced a dynamic RG that provides sufficiently high sensitivity even at low RG values (Avance NEO console generation). Using this console in combination with a 9.4 T NMR for studying real-time metabolomics with dissolution dynamic nuclear polarization (dDNP)[5], we wanted to find optimal RG and excitation angle values a priori *before the actual experiment* to optimize SNR, knowing the polarization and concentration of polarization of the system under investigation. Thus, we set out to investigate the SNR of our system as a function of RG.

We found a deviation between nominal and actual RG, similar to a previous reports.[7] Additionally, for some nuclei SNR was found to be non-monotonously affected by RG (**Figure 2-SNR**). This finding indicates that a default automated RG adjustment cannot adequately set RG as it is programmed towards maximizing the signal, not the SNR. The maximum SNR for X-nuclei was reached at a modest RG of 10-18, far below the maximum RG value of 101. This provides benefits, as it allows for signal quantification within the linear range of the receiver, avoiding signal compression as the signal stays far below 50% RRT for thermally polarized samples. If necessary, it further allows for stronger flip angles for excitation without sacrificing sensitivity. An optimal RG of in example, 18, keeps the signal more than 80% lower compared to maximum RG, which reduces the need to adjust the RG to avoid overflow and thus allowing for more samples to be samples with maximum SNR.

Below, we will discuss how to find the SNR as a function of RG, SNR(RG) (i), and the optimal RG and excitation angle parameters for thermal- and hyperpolarization experiments (ii). We exemplify the advantages of calibrating the RG by recording hyperpolarized [1-$^{13}$C]pyruvate (iii), and comparing SNR(RG) between different spectrometers (iv). The calibration procedure does not require much time and offers a considerable gain in SNR *for free*: no additional scan time is needed, and the optimal RG value can be estimated based on the sample composition (labeling, number of nuclei, concentration) and SNR(RG) calibration.

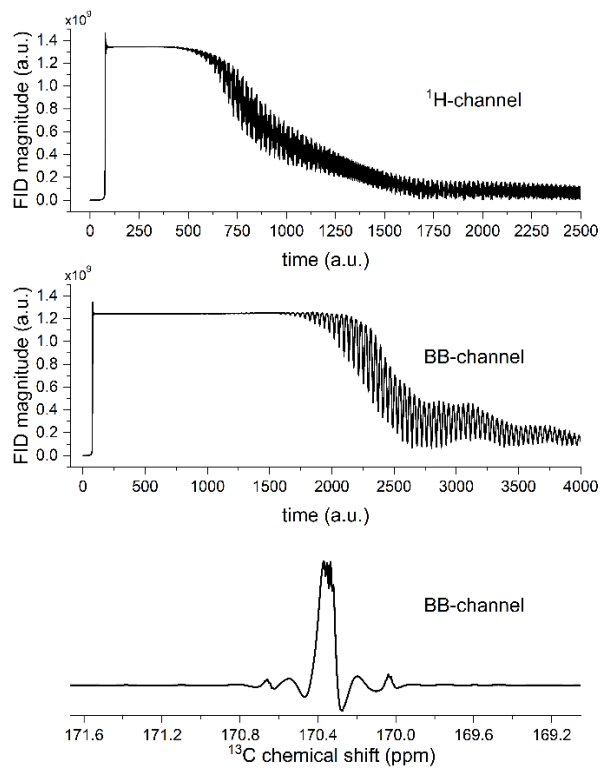

*Figure 1: Magnitude $^1$H (top) and $^{13}$C (middle) free induction decays of thermally polarized water sample and hyperpolarized [1-$^{13}$C]pyruvate (99 mM), and spectrum of the latter (bottom). The signals above ~1.3·10$^9$ a.u. were recorded incorrectly; the receiver range threshold (RRT) value causes severe distortions in the spectrum (bottom). A reliable signal quantification is not possible in this case. Signal plotted as recorded*

*by standard NMR spectrometer in the rotation frame of the excitation pulse; oscillations are caused by off-resonance excitation.*

# Materials and Methods

**NMR instrumentation**

**9.4 T NMR:** Most experiments were performed on a high resolution, 9.4 T wide bore NMR (WB400, Avance NEO, Bruker) with a 5-mm narrow bore broadband fluorine observe (BBFO) probe and 550 µL samples in standard 5 mm economy NMR tubes. The system is equipped with AV4 Transceiver (TRX) 1200 Z148391/04928, used for $^1$H measurements and TRX 1200 Z148391/04917 used for X-nuclei measurements (BB-channel). The engineering change level (ECL) determines the hardware revision, it is 02.03 for both TRX systems. The results were compared with three other Avance NEO systems. The year of installation was 2020.

**7 T MRI**: BioSpec 70/30 (Avance NEO, Bruker) with quadrature, volume, transmit-receive $^1$H (86 mm inner diameter, Bruker) and $^{13}$C imaging coils (32 mm inner diameter, Rapid Biomedical) or $^{15}$N surface coil (Rapid Biomedical); $^1$H TRX 1200 Z148391/03627 (ECL 02.01), BB TRX 1200 Z148391/03530 (ECL 02.01), installed in 2020.

**11.7 T NMR:** narrow bore NMR (NB500, Avance NEO, Bruker) with a 5-mm narrow bore broadband fluorine observe (BBFO); $^1$H TRX 1200 Z148391/00276 (ECL 01.03), BB TRX 1200 Z148391/00277 (ECL 01.03), installed in 2017.

**14.1 T NMR:** narrow bore NMR (NB600, Avance NEO, Bruker) with a 5-mm narrow bore CP-TCI H/C/N probehead; $^1$H TRX 1200 Z148391/11424 (ECL 03.02), BB TRX 1200 Z148391/11425 (ECL 03.02), installed in 2024.

**1 T benchtop.** $^{13}$C benchtop NMR with a built-in solenoid transmit-receive coil (SpinSolve Carbon, Magritek).

**Measurement of signal and SNR as a function of RG**

The NMR signal amplitude is affected, among other factors, by the following parameters[9]:

$$\text{signal} = \text{signal}(RG, \dots) = A \cdot f(RG) \cdot \sin(\alpha) \cdot P \cdot C \quad \text{eq. 1}$$

$f(RG)$ is the receiver gain function which is equal to $RG$ for a linear receiver. $\alpha$ is the angle of the excitation pulse, $P$ is the polarization value of nuclear spin of interest, $C$ is the spin concentration. $A$ is a hardware coefficient, which is independent of $RG$, $\alpha$, $P$ and $C$. A is easily determined experimentally.

Each ADC system has a maximum signal $S_m$ that can be accurately recorded, and the NMR signal should be less than this threshold:

$$\text{signal} \leq S_m \quad \text{eq. 2}$$

For example, $S_m$ could be set to 50% RRT instead of 100% RRT to avoid signal compression.[10] The noise level is a nontrivial function of many parameters, including the $RG$ value:

$$\text{noise} = \text{noise}(RG, \dots) \quad \text{eq. 3}$$

Then, the signal-to-noise ratio (SNR) can be estimated as

$$\text{SNR}(RG) = \frac{\text{signal}(RG,\dots)}{\text{noise}(RG,\dots)} = \frac{A \cdot RG \cdot \sin(\alpha) \cdot P \cdot C}{\text{noise}(RG,\dots)} \quad \text{eq. 4}$$

Grouping the known and unknown factors of eq. 4 results in:

$$\text{SNR}(RG) = \text{SNR}_{\text{ref}}(RG) \cdot \frac{\sin(\alpha)}{\sin(\alpha_{\text{ref}})} \cdot \frac{P}{P_{\text{ref}}} \cdot \frac{C}{C_{\text{ref}}}, \quad \text{eq. 5}$$

where SNR$_{\text{ref}}$(RG) includes the unknown parameters and has to be determined experimentally. Once measured, eq. 5 provides an estimate of the $SNR(RG)$ for a given set of nuclei, $RG, \alpha, P, C$.

We used the following protocol to measure $\text{SNR}_{\text{ref}}(RG)$ (**Figure 2**):

1. Place the NMR tube with 550 µL of calibration sample into the NMR probe and wait for temperature equilibration.
2. Adjust the probe resonance (match, tune), shims, and RF power.
3. Acquire FIDs as a function of $RG$ using a standard $\alpha$-pulse-acquisition protocol ($\alpha$-FID, where $\alpha$, $P$, $C$ are known). Ensure the time between experiments is at least 3×T$_1$ (consider adding Gd).

The signals were quantified by numerical integration of manually selected peaks, and the noise was calculated by taking the root mean square deviation of an automatically selected, signal-free region of the spectrum (MestReNova, version 14.2.0-26256, 2020, "*SNR Graph*"). This protocol was used to measure the reference signal and SNR (**Figure 2**).

## Calculation of optimal RG

It is a classical, constraint-satisfaction problem to find the maximum for SNR (eq. 5) as a function of many parameters with constraints. In the experiment, however, there are essentially only two parameters that change nonlinearly: $RG$ and $\alpha$. The rest are predetermined or can be treated as semifixed.

In this case, the complete optimization task can be written as:

$$\begin{cases} \text{SNR}(RG) \to max \\ \text{signal} \leq S_\text{m} \\ 0 < \alpha \leq \alpha_m \leq 90° \end{cases} \quad \text{eq. 6}$$

These conditions ensure that the signal is below the threshold signal value, $S_\text{m}$, so the ADC does not overflow (**Figure 1**, condition 2). In addition, one can put supplementary constraints on the excitation angle (condition 3): this is especially important in hyperpolarization experiments where the signal is measured using small $\alpha$. Hence, using eq. 6, one can maximize SNR for given $P$, $C$, and maximum allowed $\alpha_m$ for the experiment.

As $P$ and $C$ are typically predefined, $\alpha$ and $RG$ are the only two variables. With their variation within the constraints of eq. 6 the SNR can quickly be maximized numerically. We propose the following approach, which is visually easy to understand. The algorithm for our system is realized using a Python script and is available as supplementary material.

The approach consists of the following five steps (**Figure 3**):

1. Experimentally measure $\text{SNR}_\text{ref}(RG)$ for constant $\alpha_\text{ref}$, $P_\text{ref}$, $C_\text{ref}$ (we used the thermally polarized reference samples as described above).
2. Specify required $S_\text{m}$ and $\alpha_m$.
3. Calculate $\text{SNR}(RG, \alpha)$ and $\text{signal}(RG, \alpha)$ for $0 \leq \alpha \leq \alpha_m$ and the available RG range (for our Avance NEO, it is from 0.25 to 101)
4. Find the area where $\text{signal}(RG, \alpha) \leq S_\text{m}$.
5. Apply the mask for $\text{SNR}(RG, \alpha)$ and find the maximum.

As an alternative to the numerical approach, one can use the Karush–Kuhn–Tucker method for finding the function's maximum under the given constraints.

## NMR samples

**$^1$H SNR calibration sample No. 1:** 9.93 mg H$_2$O d.i., 580.87 mg D$_2$O (00506, Deutero GmbH), and 13.34 mg (10 uL) medical-grade Gadolinium-based contrast agent (GdCA, Gadobutrol, Gd-DO3A-butrol in water, 1 mmol/mL; Gadovist, Bayer). 10 uL GdCA contained about 7.29 mg of H$_2$O, resulting in 17.17 μL H$_2$O and 525.91 μL D$_2$O (or 1.728 M H$_2$O and a 3.456 M $^1$H concentration) in total.

**$^1$H SNR calibration sample No. 2:** 15 mL 90% H$_2$O d.i. and 10% D$_2$O.

**$^2$H SNR calibration sample No. 1:** 527.28 mg H$_2$O d.i., 11.09 mg D$_2$O, and 13.64 mg (10 μL) GdCA. This converts to 528.87 μL H$_2$O, 10.04 μL D$_2$O, and 10 μL GdCA. This leads to a 1.009 M D$_2$O and 2.018 M $^2$H concentration.

**$^{13}$C SNR calibration sample No. 1:** 400 μL H$_2$O d.i., 100 μL D$_2$O, and 50 μL [1-$^{13}$C]pyruvate (>99% $^{13}$C enrichment, 677175, Sigma-Aldrich). This leads to a 1.308 M [1-$^{13}$C]pyruvate and 1.308 M $^{13}$C concentration.

**$^{13}$C SNR calibration sample No. 2:** 13.21 mg (10 μL) GdCA and a 3.85 M [1-$^{13}$C]pyruvate in 550 μL aqueous solution (99.9 μL D$_2$O).

**$^{13}$C SNR calibration sample No. 3:** 1.2 M [$^{13}$C]urea with 50 μL GdCA in 15 mL H$_2$O d.i..

**$^{15}$N SNR calibration sample No. 1:** 494 mg H$_2$O d.i., 55.07 mg D$_2$O, 55.08 mg $^{15}$NH$_4$ (98% $^{15}$N enrichment, 299251, Sigma-Aldrich), and 10 μL Gd. This converts to 495.5 μL H$_2$O and 49.87 μL D$_2$O. This leads to a 2.059 M $^{15}$NH$_4$ and 2.059 M $^{15}$N concentration.

**$^{15}$N SNR calibration sample No. 2:** 1 M $^{15}$NH$_4$ and 19 μL GdCa in 15 mL H$_2$O.

**$^{15}$N SNR calibration sample No. 3:** 5 M $^{15}$NH$_4$ and 5 μL GdCa in 550 μL D$_2$O.

## NMR sequence parameters

**9.4 T:** For measurements at the 9.4 T system, we used $^1$H, $^2$H, $^{13}$C, and $^{15}$N samples No. 1 as described below with the following parameters. Repetition time (TR) was 30 s for $^1$H, 20 s for $^2$H, 300 s for $^{13}$C, and 80 s for $^{15}$N (e.g., T$_1^{15N}$ = 2.1 s). We used 2 averages for $^{13}$C, 8 for $^1$H and $^2$H, and 16 for $^{15}$N, and 1 dummy scan for $^{13}$C or 4 for $^1$H, $^2$H, and $^{15}$N.

**1 T benchtop:** For experiments at 1 T, we used the $^1$H reference sample No. 1 and a $^{13}$C reference sample No. 2. For $^1$H, a 90° pulse with TR of 40 s and 4 averages, and for $^{13}$C, a 90° pulse with TR of 40 s and 40 averages were used. In contrast to the Avance NEO systems, the RG of the 1 T NMR spectrometer is given in dB, and the signal is divided by the number of averages and saved to make it constant independent of the RG value. The minimum RG at 1 T is -20 dB, and the maximum is 70 dB.

**7 T MRI:** For experiments at 7 T, we used the $^1$H reference sample No. 2, $^{13}$C reference sample No. 3, and $^{15}$N reference sample No. 2. For $^1$H, a 0.176° (90°/512) pulse with TR of 20 s, 2 dummy scans, and 8 averages, for $^{13}$C, a 90° pulse with TR of 100 s, 4 dummy scans, and 16 averages was used, and for $^{15}$N, a 90° pulse with TR of 32 s, 2 dummy scans, and 32 scans was used.

**11.7 T NMR:** For experiments at 11.7 T, we used the $^1$H reference sample No. 1, the $^{13}$C reference No. 1, and the $^{15}$N reference sample No. 3. For $^1$H, a 90° pulse with TR of 10 s (>>$T_1$), 4 dummy scans, and 8 averages, for $^{13}$C, a 90° pulse with TR of 300 s, 0 dummy scans, and 1 scan was used, and for $^{15}$N, a 90° pulse with TR of 5 s, 0 dummy scans, and 1 scan was used.

**14.1 T NMR:** For experiments at 14.1 T, we used the $^1$H reference sample No. 1, the $^{13}$C reference No. 2, and the $^{15}$N reference sample No. 3. For $^1$H, a 5° pulse with TR of 5 s (>>$T_1$), 0 dummy scans, and 1 scan, for $^{13}$C, a 90° pulse with TR of 20 s, 0 dummy scans, and 1 scan was used, and for $^{15}$N, a 90° pulse with TR of 5 s, 0 dummy scans, and 1 scan was used.

In all cases, the data was processed without apodization or other filtering, due to their impact on observed signal and SNR. Only phase and baseline corrections were applied.

**dDNP experiments**

Following an established protocol[11], the pyruvate stock sample was prepared by mixing about 25 mg of trityl radical (AH111501, Polarize) and [1-$^{13}$C]pyruvate (>99% enrichment, 677175, Sigma-Aldrich). This resulted in 31 mM trityl and 14 M pyruvate concentrations in the stock sample, and 20 µL were used for hyperpolarization. After dissolution with 3.7 mL of superheated (~200°C, 11 bar) dissolution medium, the sample was transferred to an NMR system and detected after about 18 s.

Using such polarized pyruvate, we carried out two different experiments. First, we measured the maximum signal $S_m$ of the 9.4 T BB channel which can be recorded accurately. A hyperpolarized pyruvate sample (38% at the measurement site) was placed in the NMR, and a signal generated with a 40° pulse and maximum RG = 101 was used to saturate the ADC. This approach resulted in signal clipping (**Figure 1**).

In the second experiment, we measured the NMR signal at 9.4 T of a given sample repeatedly every TR = 7-10 s using $\alpha = 5°$, while alternating RG between 0.25 and 18 for every acquisition (**Figure 7**). The repetition time between experiments was accurately recorded but not constant because of the internal delay of the console needed to compile and execute individual experiments. The polarization at the measuring site was 32.9%.

**Determination of RRT at 9.4 T**

To find the ADC threshold, we measured the NMR signals of pure H$_2$O ($^1$H channel) with $\alpha = 90°$ or hyperpolarized [1-$^{13}$C]pyruvate (BB channel) with $\alpha = 40°$ at maximum $RG = 101$ and observed the maximum signal in the FID, indicating the threshold of the spectrometer's ADC (**Figure 1**). The maximum value for $^1$H was $RRT \sim 1.34 \cdot 10^9$ and for $^{13}$C $RRT \sim 1.24 \cdot 10^9$. For simplicity, we assume $RRT = S_m$ in the following. Note, that the RRT depends on the system and its TRX generation, thus the RRT must be checked for each TRX device.

## Results.
**SNR as a function of RG: detailed measurements at 9.4 T system**

To find the actual performance of the 9.4 T NMR system, we acquired thermally polarized NMR spectra for four reference samples (No. 1s) as a function of RG (**Figure 2**). Each sample resulted in a dominant, intense NMR resonance.

As expected, for all nuclei, the signal divided by sample concentration increased approximately linearly with RG with a slope (units a.u./(RG·M)) of ≈ $1.21 \cdot 10^6$ for $^1$H, ≈ $2.27 \cdot 10^5$ for $^2$H, ≈ $2.84 \cdot 10^4$ for

$^{13}$C, ≈ 3.65·10$^3$ for $^{15}$N. The SNR per concentration showed an asymptotic increase to a maximum value, ca. 2009 1/M for $^1$H, 6584 1/M for $^2$H, 65 1/M for $^{13}$C, and 59 1/M for $^{15}$N (90° excitation).

To evaluate the interplay of RG and signal further, we divided the ratio of signal per RG, signal(RG)/RG, by the same ratio at maximum RG, signal(RG = 101)/101, which should always give one if the RG is linear:

$$\text{signal}_{\text{deviation}}(RG) + 1 = \frac{\text{signal}(RG)}{RG} \cdot \frac{RG_{\max}}{\text{signal}_{\max}}$$

eq. 7

We found that $^1$H and $^2$H signals are fairly linear concerning RG (maximum signal$_{\text{deviation}}$ = 1-2%), while $^{13}$C and $^{15}$N display substantial deviations in linearity between (up to 8% and 30%, respectively, **Figure 2**, right). Despite the signal and SNR of $^{13}$C and $^{15}$N being about 100 times lower compared to $^1$H and $^2$H, the SNR for $^{13}$C and $^{15}$N is about 50, even for the lowest RG. At this SNR, noise variations are too small to account for the observed strong single variations. However, it may suggest that lower signals are more prone to experience RG non-linearity, which should be checked experimentally. Note that in all cases, the signal was below RRT of $1.3 \cdot 10^9$. In cases of X-nuclei, the signal was at least 10 times lower than the RRT, excluding signal compression as another potential cause for signal non-linearity.

Unexpectedly, we found a strong dip in SNR at 9.4 T of about 40% for all measured nuclei except $^1$H at $RG$ =18. As the signal increases linearly, this effect is caused by noise intensity. The plateau was reached at $RG$ > 30 for $^1$H and $10 \leq RG \leq 18$ for the other nuclei.

These findings showed that the most sensitive RG for X-nuclei is between 10 and 18 and should be used in most cases without any automatic RG adjustment. Lower RG values are needed when the signal exceeds the threshold, while higher RG values do not provide any benefits. The $^1$H channel showed the most gain in SNR up to $RG \approx 30$, with minor improvements thereafter but with a potential threat of signal compression. This finding revealed that choosing a bigger RG is not necessarily the right strategy to improve SNR.

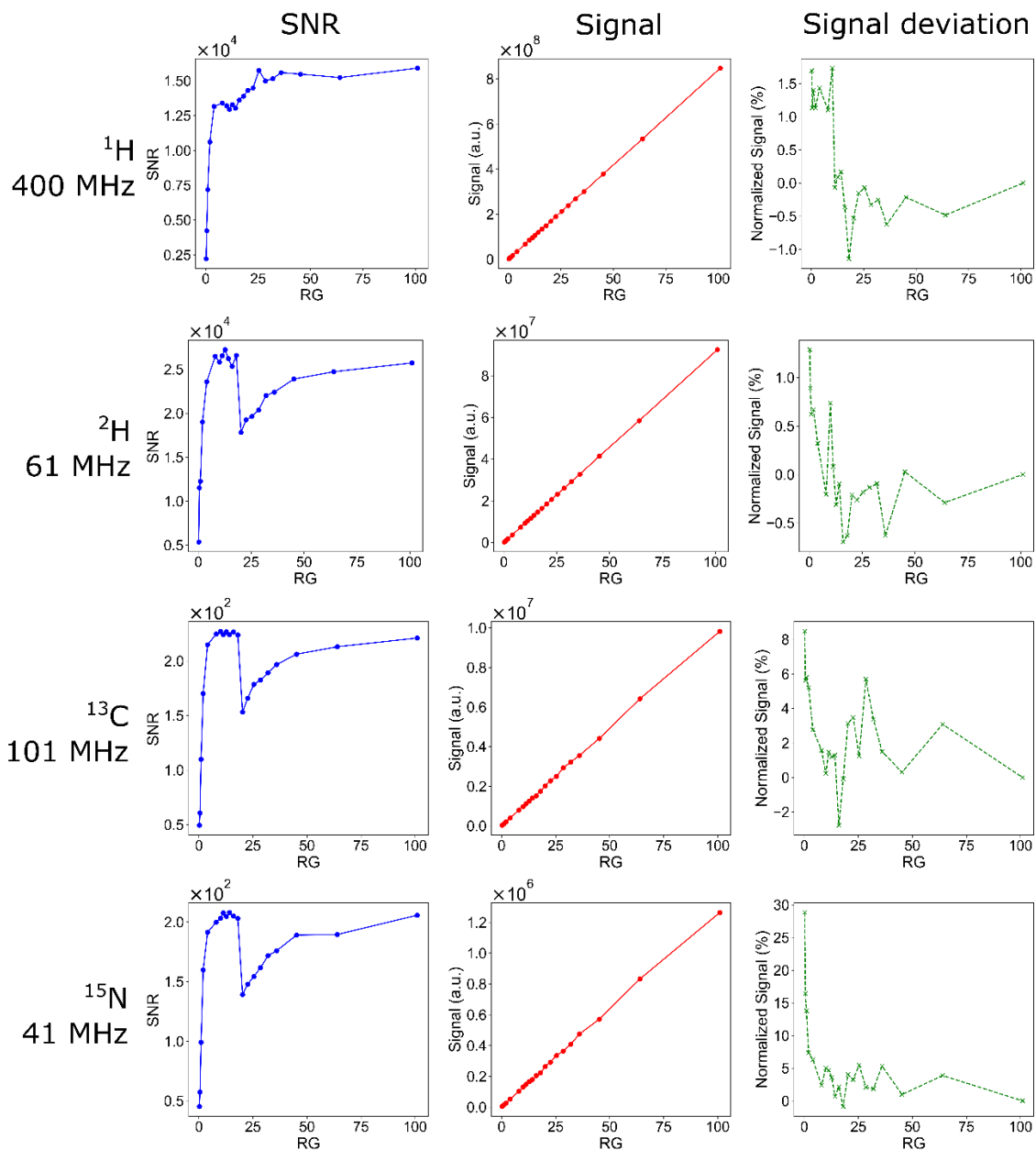

*Figure 2: SNR (left), signal (middle), and deviation of the signal from signal(RG = 101)/101 (right) for $^1H$, $^2H$, $^{13}C$, and $^{15}N$ reference samples as a function of RG at 9.4 T at 90° excitation. The signal was found to increase linearly with RG, as expected, for all nuclei (middle). The relative deviation of the signal(RG)/RG to signal(RG = 101)/101 (right) showed larger variations for RGs below 20, and larger deviations for $^{13}C$, $^{15}N$ than for $^1H$ and $^2H$. The SNR approached a maximum value asymptotically, as expected, but showed an unexpected, abrupt drop at RG 18 for all nuclei but $^1H$. After the drop, the SNR increased again asymptotically but slower than before.*

**Optimal settings for a hyperpolarization observation**

Having measured $\mathrm{SNR}(RG)_{\mathrm{ref}}$ (**Figure 2**) and the RRT ($S_m \sim 1.34 \cdot 10^9$ for $^1H$ and $S_m \sim 1.24 \cdot 10^9$ for other nuclei at the 9.4 T system, **Figure 1**), one can estimate the signal intensity and SNR for a known sample for any parameters: $RG$, $\alpha$, $C$, and $P$.

Using eq. 1 and 5, we calculated the signal and SNR for pure water (55.51 M) and D$_2$O (55.21 M) for $\alpha = 0 - 90°$ and RG = 0 − 101 (**Figure 3**. Note that the concentration of the $^1$H and $^2$H spins is twice as large as the concentration of the molecules). For more than half of the combinations of $\alpha$ and RG, the signal exceeded S$_m$ (white areas in **Figure 3**, right). These results indicate that $\alpha$ and $RG$ should be chosen with care.

Note that if there are no constraints on the flip angle, maximizing $\alpha$ is more beneficial for SNR than maximizing RG (**Figure S2**, SI). For water, the maximum SNR was obtained with $\alpha = 90°$ and $RG$ of 4.9 ($^1$H) and 12.8 ($^2$H) – values far below the maximum RG of 101.

In hyperpolarization experiments, $C$ and $P$ are usually given, while $\alpha$ and $RG$ can be adjusted as needed. We calculated the signal and SNR as a function of $RG$ and $\alpha$ for a 90 mM $^{13}$C sample polarized to 35% and a 40 mM $^{15}$N sample polarized to 15% - similar values were achieved recently with [1-$^{15}$N]nicotinamide[12] (**Figure 3**). To preserve polarization, the flip angle $\alpha$ in hyperpolarized metabolic experiments is typically low ($\sin(\alpha) \approx \alpha$). Here, we set the limits to 5° ($^{13}$C) and 10° ($^{15}$N).

For the given conditions, we calculated the maximum product of $\sin(\alpha) \cdot RG$ that can be used without clipping to 4.99 for $^1$H, 24.82 for $^2$H, 4.14 for $^{13}$C, and 107.15 for $^{15}$N. This means that, e.g., for hyperpolarized $^{13}$C with a 90° pulse, only $RG \leq 4.14$ avoids clipping, while for $^{15}$N, any combination of RG (from 0.25 to 101) and flipping angle (from 0 to 90°) can be used.

When the nonlinear behavior of $SNR(RG)$ is also considered (**Figure 2**), the maximum SNR was obtained for all X-nuclei and $\alpha < 5°$ at $10 \leq RG \leq 18$, where $SNR(RG)$ reached its maximum before going down.

Note that in this work, we use an exact condition for the signal overflow: $\text{signal} \geq \text{RRT}$. Routinely, it is practical and safer to use a more conservative overflow definition of $\text{signal} \geq 0.5 \cdot \text{RRT}$ that leaves some room for errors without actual signal overflow[10].

Typical metabolic hyperpolarization studies are conducted with $^{13}$C labeled substrates. When the hyperpolarized tracer is injected into cell cultures, the bloodstream or similar, dilution occurs so that maximum SNR will be obtained at RG = 18 with little risk of clipping (assuming typical polarization levels of 35 %). Raising the RG beyond this value provides no SNR gains.

For the example of a solution in **Figure 3** (90 mM $^{13}$C-pyruvate solution and 35% polarization), the best flip angle (providing highest SNR) is about 13° at RG 18, or 27° at 45 mM after e.g. 1:1 dilution with cells, or 90° at much higher dilution in the blood.

**Experimental evidence that hyperpolarization experiments benefit from RG calibration**

To avoid variations between different hyperpolarization experiments, we investigated the effect of varying RGs in one single experiment, where we alternated the receiver gain between 0.25 and 18 ($\alpha = 5°$, TR = 7 – 10 s, **Figure 4**). We used 100 mM [1-$^{13}$C]pyruvate polarized to ≈ 33% (at the time of detection) and observed a five-fold increase in SNR for RG = 18 (compared to RG = 0.25), impressively demonstrating the gain in sensitivity.

The signal was normalized using the $^{13}$C calibration experiment (**Figure 2**) to obtain the ratio of signals at RG 0.25 and 18 (66.33 experimentally, instead of 18/0.25 = 72 expected for perfectly linear RG). Following normalization, the mono-exponential decay inherent to hyperpolarized signals is apparent despite the alternating RGs.

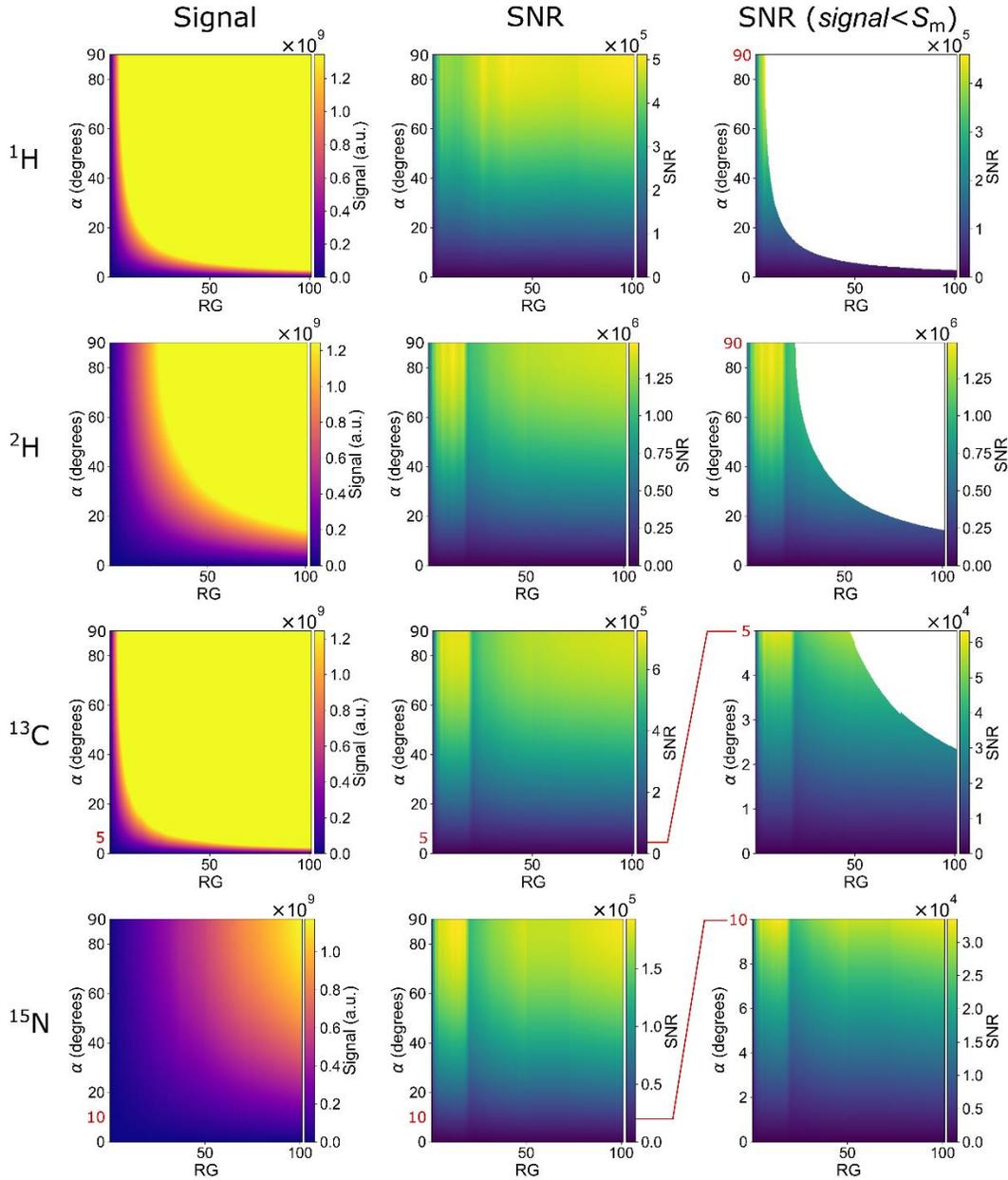

*Figure 3: Calculated 9.4 T NMR signals (left) and SNR, including (center) and excluding (right) parameters where clipping is expected, as a function of excitation angle α and RG for H₂O (111.02 M ¹H) and D₂O (110.42 M ²H), both thermally polarized, or 90 mM, 35% hyperpolarized ¹³C and 40 mM, 15% hyperpolarized ¹⁵N.* Note the continuous, monotonous variation of the signal, and the discrete jumps for X-nuclei SNR. Signals exceeding the maximum detectable signal $S_{max}$ were removed from the right column. The data was calculated using eq. 5. The following parameters were used in simulations, which should reflect possible experimental conditions: (¹H) a pure H₂O sample at thermal polarization was assumed with [H₂O] = 55.51 M, (²H) a pure D₂O sample with [D₂O] = 55.21 M, (¹³C) a 90 mM ¹³C with 35% polarization, and (¹⁵N) a 40 mM ¹⁵N with 15% polarization. The maps were calculated by extrapolating signal(RG) and SNR(RG) reference maps (*Figure 2*) for missing RG values between 0.25 and 101. Maximum flip angle was limited to 5° (¹³C) and 10° (¹⁵N) to mimic experimental conditions following hyperpolarization. Values when $\text{signal} > S_m$ were removed from the SNR maps (third column, white areas), leaving behind only allowed ($\text{signal} < S_m$, when ADC is not overflown) combinations of α and RG.

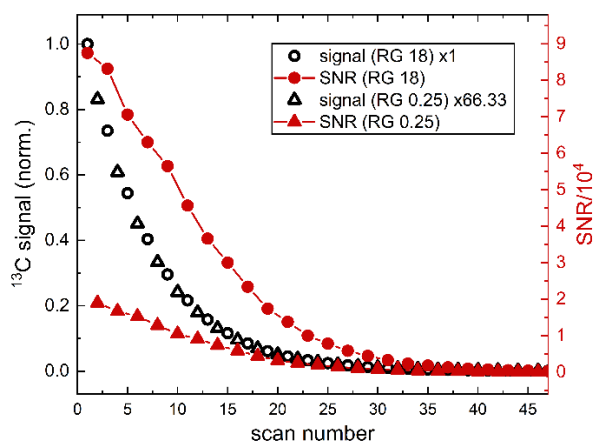

*Figure 4: $^{13}C$ NMR integrals (black) and SNR (red) of hyperpolarized [1-$^{13}C$]pyruvate acquired 46 consecutive spectra after α = 5° excitations with alternating RG (0.25 - triangles / 18 - circles). The integrals of RG 0.25 were multiplied by calibrated factor 66.33 (**Figure 2**) to bring it to the same level as the signal obtained with RG 18. Both signal and SNR were found to decay monotonously, as if measured with the same RG. The SNR was about 5 times higher for RG 18 than RG 0.25, similar to the results of the calibration experiments (factor 4.8). The repetition time was 7 to 10 s because of compilation delays in the console.*

**SNR as a function of RG: intersystem comparison**

We also tested the matter on 7 T MRI, 11.7 T NMR, and 14.1 T NMR systems of the same manufacturer with Avance NEO consoles (**Figures 4** and **5**). Interestingly, at 7 T, we found similar signal and SNR dependences on RG for $^1H$ and $^{13}C$ (**Figure S3**). This finding indicates that our results at 9.4 T are likely not the consequence of faulty equipment but rather a feature of the hardware. Note, however, that the drop in SNR occurred after RG 16 at the 7 T system instead of RG 18 at 9.4 T.

In comparison, the results at 11.7 T show a steep increase of both $^1H$ and $^{13}C$ up to RG of about 10 and a plateau thereafter (**Figure S4**). At 14.1 T, the increase of $^1H$ and $^{13}C$ SNR is less steep and is monotonously growing up to 101 with a declining growth rate of SNR (**Figure S5**). The different behavior in SNR can be easily observed (**Figure 5**).

Notably, for $^{15}N$, the behavior of the 11.7 T and 14.1 T systems changes, similar to that of the 7 T and 9.4 T systems. The SNR was found to drop after RG 4 at 11.7 T by 70% and after RG 28.5 at 14.1 T by 23% (**Figure 6**).

Repeating the experiments on a 1 T NMR spectrometer of a different manufacturer did not show a drop in the SNR(RG), but unexpected behavior of the signal (SpinSolve, Magritek, **Figure 7**). The signal showed a monotonously growing trend, increasing by about 50% from RG = -20 dB to 70 dB, and some variation, e.g., for high and low RG. Theoretically, the spectrometer is supposed to correct for RG automatically and thus deliver constant signals (as per manual), which was not the case. The SNR grew slowly to ≈ 0 dB, then rapidly leveled at > 25 dB. This finding may suggest an imperfect signal(RG) calibration that the user should always check.

Note, that Bruker systems can be configured to produce a signal independent of RG or NS, similar to the system of Magritek, by selecting the according acquisition settings in the preferences.

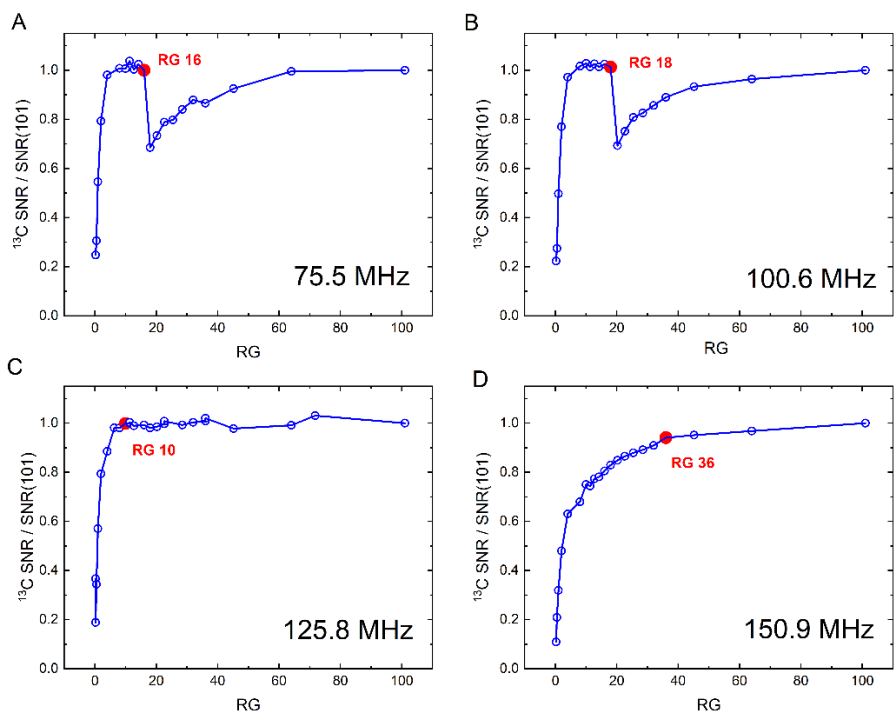

*Figure 5: $^{13}$C SNR graphs at different field strengths of 7 (A), 9.4 (B), 11.7 (C), and 14.1 T (D) Bruker systems with Avance NEO console.* At 7 and 9.4 T, a steep increase up to RG 16 and 18, respectively, can be observed, with a drop and prolonged recovery of SNR thereafter. At 11.7 T, the SNR increases rapidly up to about RG 10 and plateaus thereafter. At 14.1 T, the SNR increases more rapidly at lower RG values up to RG 36 and keeps increasing monotonously until the maximum RG of 101 is reached. At the two lower field systems, the carbon-13 resonance frequency is below 101.25 MHz – the threshold value for the observed effect.

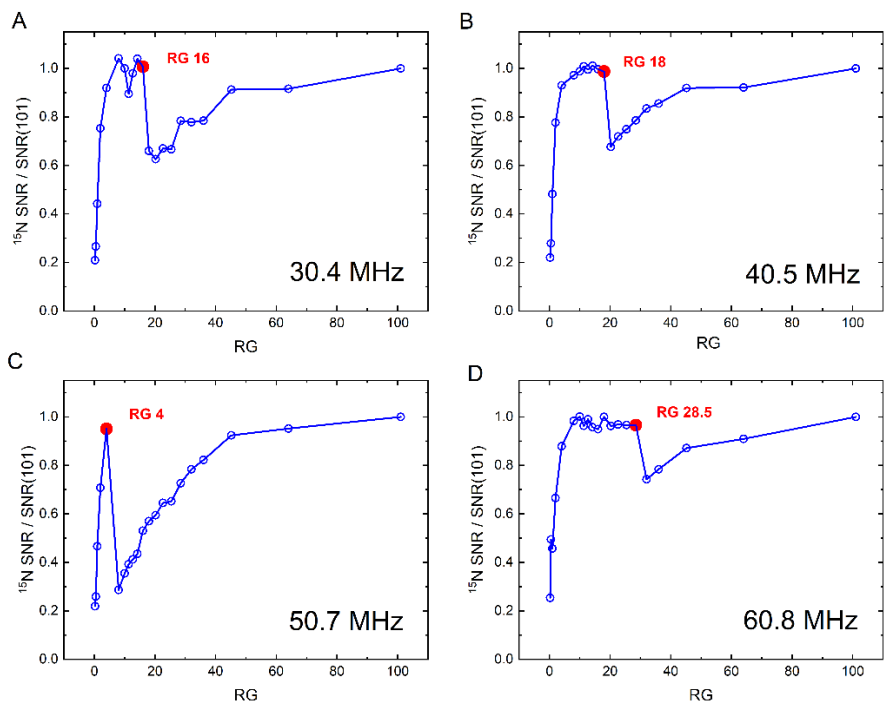

*Figure 6: $^{15}$N SNR graphs at different field strengths of 7 (A), 9.4 (B), 11.7 (C), and 14.1 T (D) Bruker systems with Avance NEO console. Similar to $^{13}$C, at 7 and 9.4 T, a steep increase up to RG 16 and 18, respectively, can be observed, with a drop of SNR and a prolonged recovery of SNR thereafter. At 11.7 T, the SNR increases rapidly up to about RG 4, drops by about 70%, and recovers thereafter. At 14.1 T, the SNR increases monotonously up to RG 28.5 and drops by about 23%, the SNR recovers until the maximum RG of 101 is reached. At all systems, the nitrogen-15 resonance frequency is below 101.25 MHz – the threshold value for the observed effect.*

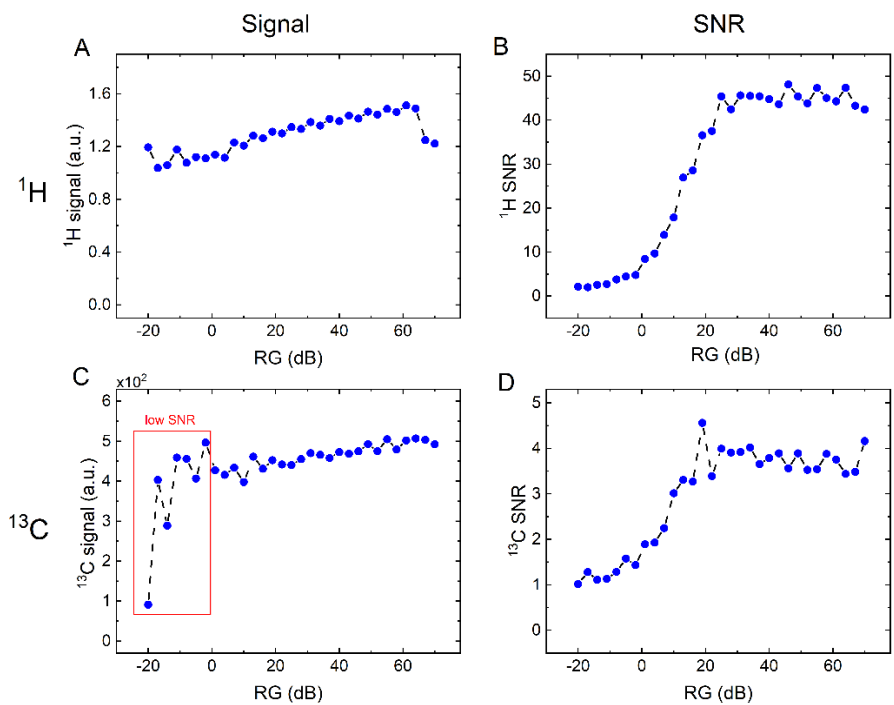

*Figure 7: $^1$H (top) and $^{13}$C (bottom) high-resolution NMR signals (left) and SNR (right) as a function of receiver gain acquired with a 90° excitation on a 1 T benchtop NMR spectrometer using the reference samples. The $^1$H and $^{13}$C signals showed a slight but steady increase with RG (e.g., 50% signal increase for $^1$H). Note that the spectrometer is supposed to correct for RG automatically (the signal should be constant). However, the signal varied markedly, e.g., for high RG. The SNR increased slowly for RG <≈ 0, then rapidly for RG ≈ 0 – 25, and leveled thereafter. The $^{13}$C SNR of about 1 for RG below ~ 0 dB is insufficient for reliable analysis.*

## Discussion

Testing and setting the RG appears to be essential for optimal NMR experiments and should not be left to automatic adjustment – at least for the devices tested here.

For example, in one of our previous studies,[13] we used RG of 10 for measuring hyperpolarized $^{13}$C signal at 9.4 T. This was a lucky pick – in the present study, we found that the SNR(RG = 10) is 1.5% higher than at SNR(RG = 18) and 2.8% higher than SNR(maximal RG = 101). However, if we had chosen an RG of 20.2 between these values, the SNR would have been about 1/3 less compared to SNR(RG = 10, 18, or 101). This is relevant for both thermally and hyperpolarized experiments. Notably, improving SNR by only 50% gives a factor of 2 acceleration when signal averaging is required. The SNR improvement of 107% from RG 1 to 10 would implicate a 4.3-fold acceleration.

Considering current advancements in hyperpolarization, levels over 60% polarization could be feasible soon[14] – using the $^{13}$C example from **Figure 3** with 70% polarization at 90 mM while maximizing RG to 101 would limit the flip angle to about 1.1°. Instead, using RG 18 would provide the same SNR from the receiver while allowing for 6 times the SNR by using a 6.6° flip angle. In this context, improved SNR performance at lower RG value, as observed on all machines, provides a benefit for hyperpolarization experiments because, despite their high polarization, e.g., metabolic experiments are limited by their low SNR.

Of course, all the metrics in this work depend on quantifying signal and SNR. Here, we used an automatic, integration-based approach. Peak fitting and manual selection of the noise may improve the quantification, thus providing even more accurate results. Still, our measurements here reveal how to stay below the FID clipping threshold and at the optimal RG during those experiments. By more careful adjustment of SNR, using 18 instead of 1, 10, 20.2, or 101 (arbitrary values once may choose), we could have changed the SNR of 9.4 T system by +104%, -1.5%, +46%, or +1.3%, respectively; specific RG ranges are very close to maximum SNR despite drastically different signal amplitudes.

We observed a similar $SNR(RG)$ profile at all field strengths for nuclei with a resonance frequency below a cut-off frequency of 101.25 MHz (see explanation below). While the 11.7 and 14.1 T machines did not exhibit such behavior for $^{13}$C due to the higher resonance frequency, they did for $^{15}$N. The actual $SNR(RG)$ figure was found to depend on the production year, field, and frequency. According to Bruker, the next TRX revision will have a different SNR to RG behavior.

The 1 T spectrometer with different hardware had no dip in the $SNR(RG)$ profile and a maximum at about 20 dB.

However, the central message of the manuscript does not depend on the specific system one uses: Although it is known that modern NMR systems are very reliable, it is still recommended to check your $SNR(RG)$ profile and subsequently setting the parameters right for your system to gain maximum SNR in hyperpolarization or conventional NMR experiments. The stability of these calibrations with time presently remains unknown and may be reported in a future study. Checking SNR(RG) regularly appears advisable.

**Statement from the manufacturer**

The manufacturer Bruker provided us with a statement to explain this phenomenon of the systems.

*"The current Bruker 'AV4 TRANSCEIVER 1200' (TRX 1200) uses different signal paths for frequencies <101.25 MHz and higher frequencies. The receiver gain is distributed over several gain stages to balance the input signal from the probe/sample/preamplifier to the optimum input level required by the ADC. A variable distribution of gain over these stages, a so-called level plan, optimizes for the highest SNR for weak signals or better dynamic range, i.e., fewer intermodulations for strong signals, leading to a non-monotonic behaviour of SNR at receiver gains ~20 in the low-frequency path. It is recommended to go as high as possible with the receiver gain but to stay below a receiver gain of 20 to get the highest SNR. The behavior depends on the hardware revisions of the TRX devices, and the software will be adapted to achieve optimal results."*

## Conclusion

The receiver gain is one of the essential parameters for NMR experiments. Erroneous settings cost SNR or cause artifacts. For experiments at thermal polarization and with solvent suppression RG can be maximized, but signal compression may occur. Using the lower RG values providing maximum SNR is safer and suggested by the manufacturer, as high RG values may behave unexpectedly amplify noise and signal equally. However, our results show that manual tests and adjustments are relevant to avoid unnecessary issues and optimize sensitivity, even on devices of the latest generation.

## Acknowledgements

We acknowledge funding from German Federal Ministry of Education and Research (BMBF) within the framework of the e:Med research and funding concept (01ZX1915C, 03WIR6208A hyperquant), DFG (555951950, 527469039, 469366436, HO-4602/2-2, HO-4602/3, HO-4602/4, EXC2167, FOR5042, TRR287). MOIN CC was founded by a grant from the European Regional Development Fund (ERDF) and the Zukunftsprogramm Wirtschaft of Schleswig-Holstein (Project no. 122-09-053).

## Abbreviations

Analog-to-digital-converter (ADC)
Broadband (BB)

Concentration (C)
Deionized (d.i.)
Dissolution dynamic nuclear polarization (dDNP)
Free induction decay (FID)
Flip-angle ($\alpha$)
Gadolinium-based contrast agent (GdCA)
Maximum flip-angle ($\alpha_m$)
Maximum signal ($S_m$)
Polarization (P)
Receiver gain (RG)
Receiver range threshold (RRT)
Repetition time (TR)
Signal-to-noise-ratio (SNR)

# Supporting Materials
Additional calibrations, FIDs, spectra, and analyzed data are available in supporting materials (.pdf).

# Data Availability
The data that support the findings of this study are openly available in Zenodo at https://doi.org/10.5281/zenodo.13795731.

# Author Information

Josh Peters, ORCID 0000-0003-1019-4067
Prof. Jan-Bernd Hövener: ORCID 0000-0001-7255-7252;
Prof. Frank Sönnichsen, ORCID 0000-0002-4539-3755;
Dr. Andrey N. Pravdivtsev, ORCID 0000-0002-8763-617X

# Author Contributions
ANP, JBH: conceptualization; ANP, JP: investigation, analysis, writing – original draft; ANP and JBH: supervision, funding acquisition. JP and AP: 42, 300 and 400 MHz investigation. JP and FS: 500 and 600 MHz investigation. All authors contributed to discussions and interpretation of the results and have approved the final version of the manuscript.

# References

1. Sapir G, Steinberg DJ, Aqeilan RI, Katz-Brull R. Real-Time Non-Invasive and Direct Determination of Lactate Dehydrogenase Activity in Cerebral Organoids—A New Method to Characterize the Metabolism of Brain Organoids? *Pharmaceuticals*. 2021;14(9):878. doi:10.3390/ph14090878

2. Sharma R, Sharma A. 21.1 Tesla Magnetic Resonance Imaging Apparatus and Image Interpretation: First Report of a Scientific Advancement. *Recent Pat Med Imaging*. 2011;1(2):89-105.

3. Kovacs H, Moskau D, Spraul M. Cryogenically cooled probes—a leap in NMR technology. *Prog Nucl Magn Reson Spectrosc*. 2005;46(2):131-155. doi:10.1016/j.pnmrs.2005.03.001

4. Hövener J, Pravdivtsev AN, Kidd B, et al. Parahydrogen-Based Hyperpolarization for Biomedicine. *Angew Chem Int Ed*. 2018;57(35):11140-11162. doi:10.1002/anie.201711842

5. Ardenkjaer-Larsen JH, Fridlund B, Gram A, et al. Increase in signal-to-noise ratio of > 10,000 times in liquid-state NMR. *Proc Natl Acad Sci*. 2003;100(18):10158-10163. doi:10.1073/pnas.1733835100

6. Eills J, Budker D, Cavagnero S, et al. Spin Hyperpolarization in Modern Magnetic Resonance. *Chem Rev*. 2023;123(4):1417-1551. doi:10.1021/acs.chemrev.2c00534

7. Mo H, Harwood JS, Raftery D. Receiver gain function: the actual NMR receiver gain. *Magn Reson Chem*. 2010;48(3):235-238. doi:10.1002/mrc.2563

8. Beckman RA, Zuiderweg ERP. Guidelines for the Use of Oversampling in Protein NMR. *J Magn Reson A*. 1995;113(2):223-231. doi:10.1006/jmra.1995.1083

9. Mo H, Harwood J, Zhang S, Xue Y, Santini R, Raftery D. R: A quantitative measure of NMR signal receiving efficiency. *J Magn Reson*. 2009;200(2):239-244. doi:10.1016/j.jmr.2009.07.004

10. Mo H, Harwood JS, Raftery D. A quick diagnostic test for NMR receiver gain compression. *Magn Reson Chem*. 2010;48(10):782-786. doi:10.1002/mrc.2662

11. Ferrari A, Peters J, Anikeeva M, et al. Performance and reproducibility of 13C and

15N hyperpolarization using a cryogen-free DNP polarizer. *Sci Rep*. 2022;12(1):11694. doi:10.1038/s41598-022-15380-7

12. Peters JP, Brahms A, Janicaud V, et al. Nitrogen-15 dynamic nuclear polarization of nicotinamide derivatives in biocompatible solutions. *Sci Adv*. 2023;9(34):eadd3643. doi:10.1126/sciadv.add3643

13. Peters JP, Assaf C, Mohamad FH, et al. Yeast Solutions and Hyperpolarization Enable Real-Time Observation of Metabolized Substrates Even at Natural Abundance. *Anal Chem*. 2024;96(43):17135-17144. doi:10.1021/acs.analchem.4c02419

14. Korchak S, Mamone S, Glöggler S. Over 50 % 1H and 13C Polarization for Generating Hyperpolarized Metabolites—A para-Hydrogen Approach. *ChemistryOpen*. 2018;7(9):672-676. doi:10.1002/open.201800086

Supporting information for:

# Maximizing NMR Sensitivity: A Guide to Receiver Gain Adjustment

Josh P. Peters[1*], Frank D. Sönnichsen[2], Jan-Bernd Hövener[1], Andrey N. Pravdivtsev[1*]

1. Section Biomedical Imaging, Molecular Imaging North Competence Center (MOIN CC), Department of Radiology and Neuroradiology, University Medical Center Kiel, Kiel University, Am Botanischen Garten 14, 24114, Kiel, Germany

2. Otto Diels Institute for Organic Chemistry, Kiel University, Otto Hahn Platz 4, 24118, Kiel, Germany

*Corresponding Authors: josh.peters@rad.uni-kiel.de; andrey.pravdivtsev@rad.uni-kiel.de

## Contents

# Signal overflow of $^{13}$C-channel at the 1 T system

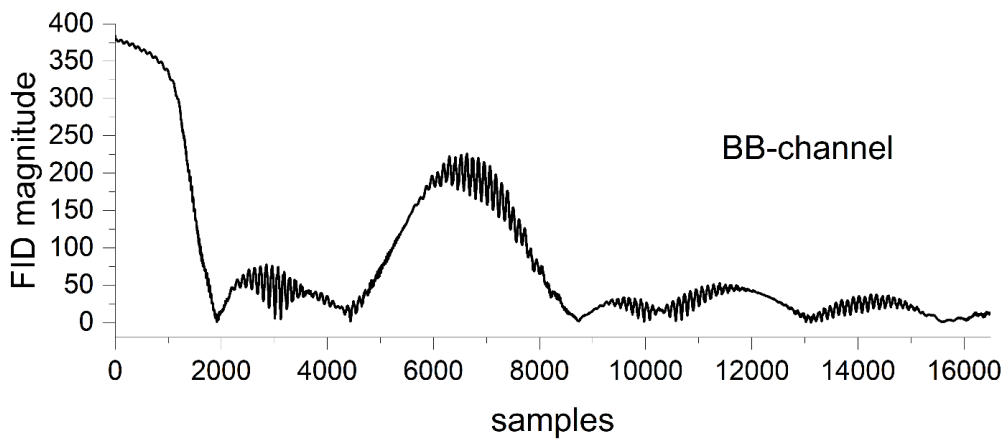

*Figure S1: Magnitude $^{13}$C (bottom) NMR free induction decay of hyperpolarized 1-$^{13}$C pyruvate (96 mM) at 1 T, where signals >≈ 380 a.u. were not adequately recorded.* Signal plotted as recorded by standard NMR spectrometer in the rotation frame of the excitation pulse; small oscillations are caused by off-resonances. The rapid decay and rising of the signal are likely caused by dynamic adjustments by the spectrometer which limits the signal to avoid overflow: This behaviour was observed in several experiments, when overflow was achieved. The signal overflow in this figure was achieved using 96 mM hyperpolarized sample with a 66° pulse and RG of 31 dB. The maximum value achieved was 384.2 a.u.

# Constant signal experiment at the 9.4 T system

**Constant signal experiment as a function of RG:** We did one experiment where we kept a constant signal(RG) = const. According to eq 1, this can be achieved by setting $RG \cdot \sin(\alpha)$ to a constant value. If one sets $\alpha = 90°$ for the lowest RG of 0.25, then

$$\sin(\alpha) = \frac{0.25}{RG} \qquad \text{eq S1}$$

gives the $\alpha$ angle as a function of RG, providing a constant signal. Using this approach, we acquired an RG sweep with varied flip angles from 90° (RG 0.25) to 0.142° (RG 101) to keep a constant signal amplitude. The highest SNR was obtained for the combination of RG 0.25 and 90° FA and subsequently fell monotonously. This indicates that for a given signal threshold, it is best first to maximize the flip angle within the constraints of the experiment and then maximize RG as a second measure. However, maximizing the flip angle is typically not feasible, especially during a hyperpolarization experiment. In this case, the optimization procedure from the main text applies.

Note that the signal decreases from $0.85 \cdot 10^5$ a.u. to $0.60 \cdot 10^5$ a.u. between RG 0.25 to 45.2, which is consistent with the finding from the RG sweeps in **Figure 2** in the main text where the signal at lower RG values is over-proportionally high.

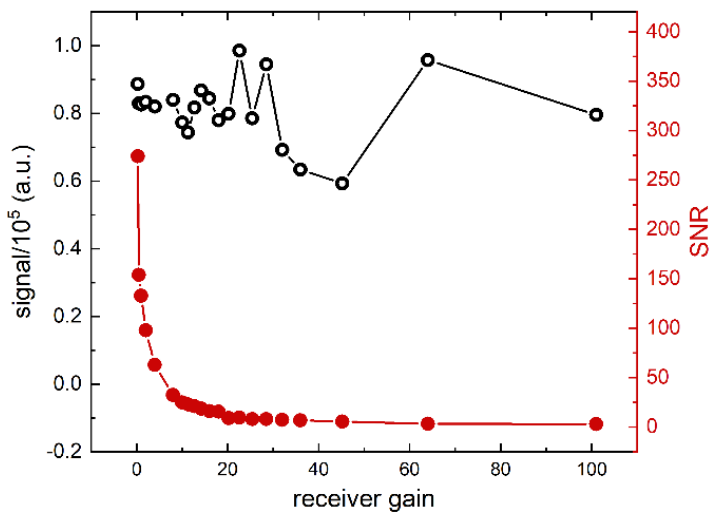

*Figure S2: Constant signal experiment on the $^{15}$N-sample at 9.4 T showcases the linearity of the signal and calibration of the 90° pulse. The signal gets noisier for higher RGs, because the flipping angle gets as low as 0.142° at RG 101. It also demonstrates that for a given threshold (here signal ≈ $0.85 \cdot 10^5$ a.u.) one should maximise first the flipping angle and then RG if maximum SNR is the goal. It is apparent, similar to Figure 3-$^{15}$N from the main text, that the signal is higher for lower RG values if sin(α)·RG is kept constant. In theory, the signal should stay constant if the RG is linear, however here, it decreases from $0.85 \cdot 10^5$ a.u. (RG 0.25) to $0.6 \cdot 10^5$ a.u. (RG 45.2), a decrease by 30%.*

# Signal and SNR sweep data at 7, 11.7, and 14.1 T

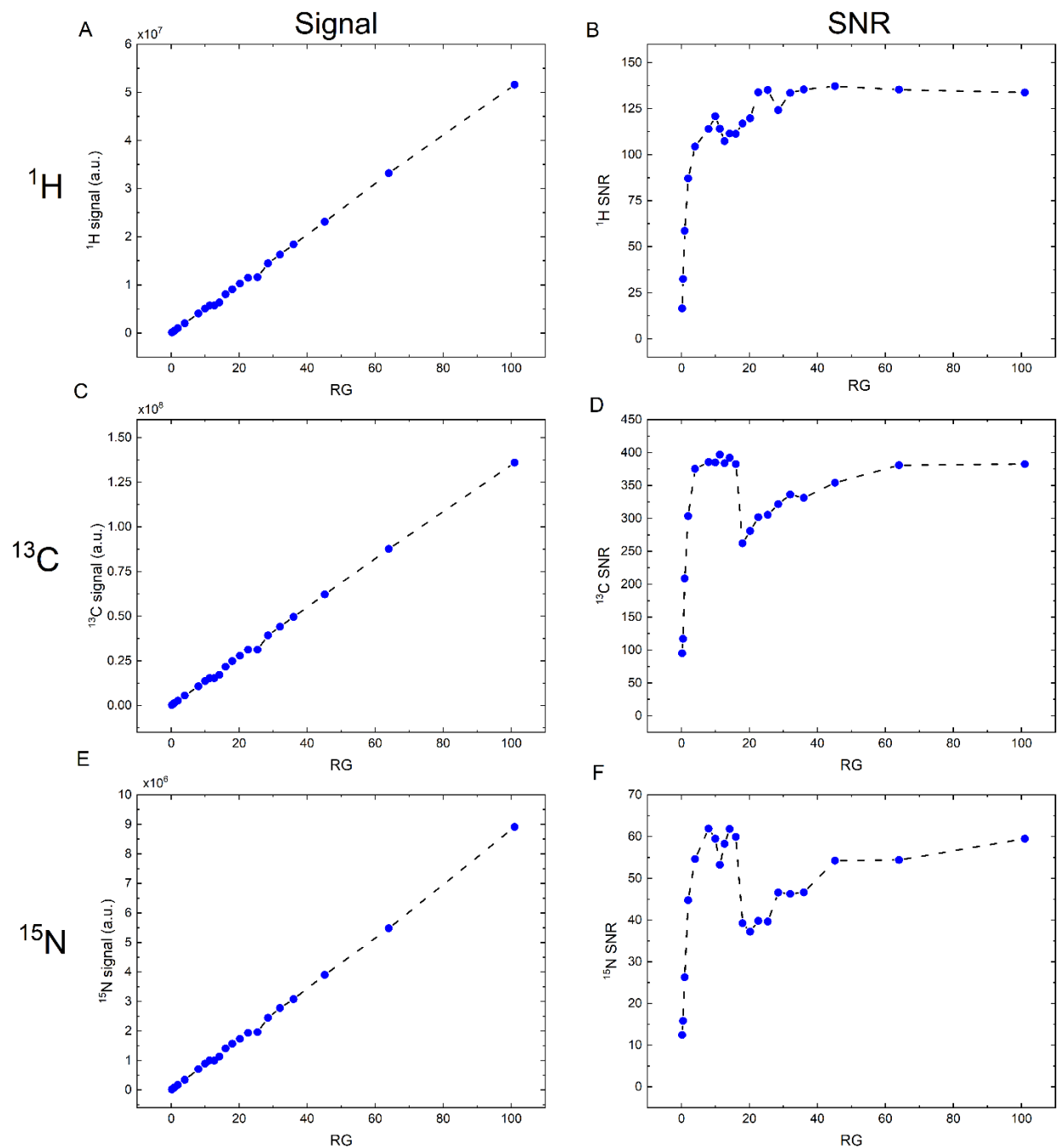

***Figure S3: Signal (left column), corresponding SNR (right column), <sup>1</sup>H and <sup>13</sup>C reference samples as a function of RG at 7 T using the Avance NEO console.*** *The reference sample composition is given in the methods section. Note the maximum and a dip in $SNR(RG)$ after $RG = 16$ and the prolonged rate of SNR recovery for the $^{13}C$ sample.*

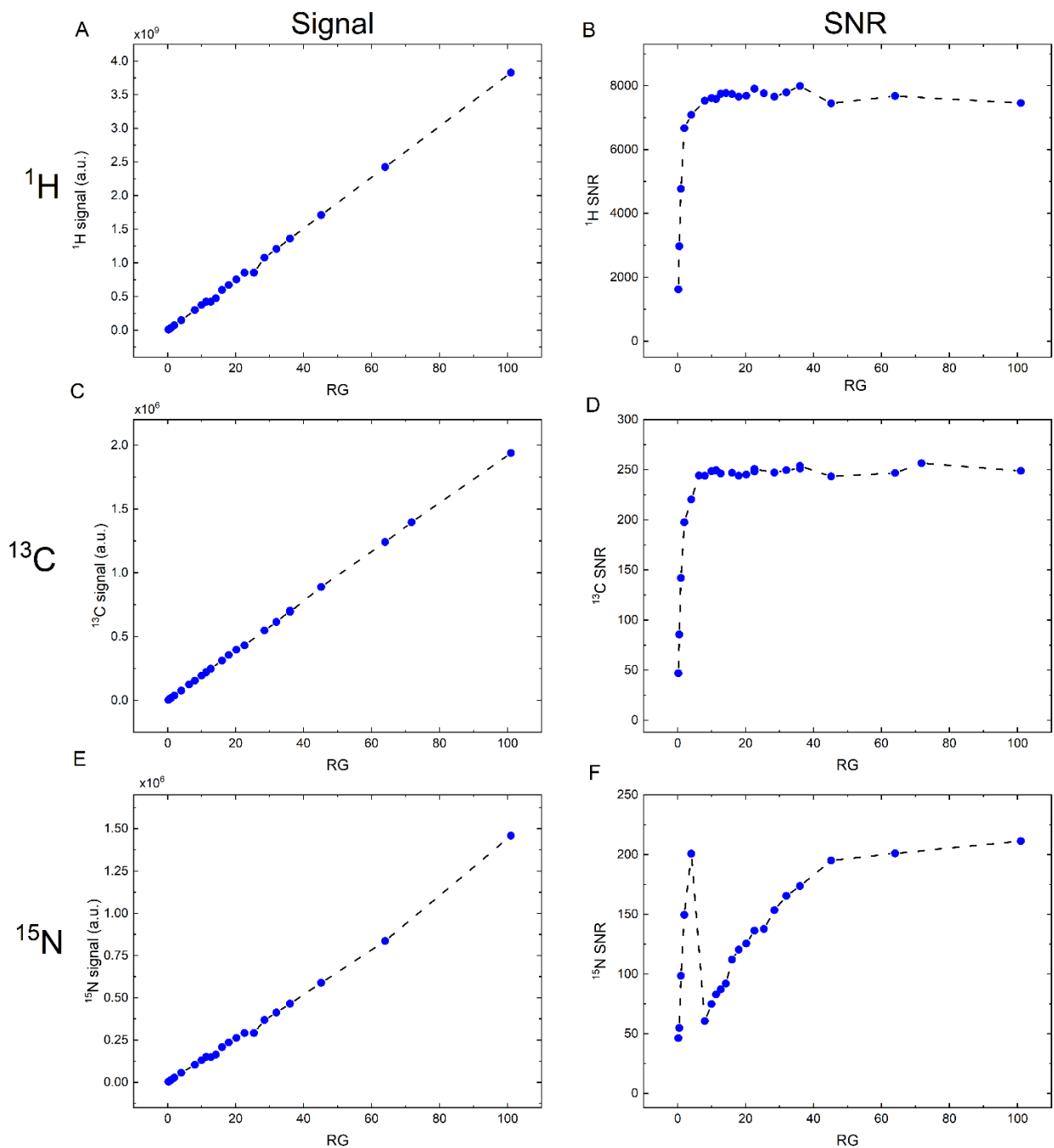

*Figure S4: Signal (left column), corresponding SNR (right column), $^1H$ and $^{13}C$ reference samples as a function of RG at 11.7 T using the Avance NEO console.* The reference sample composition is given in the methods section. Note the maximum SNR is reached around RG 10 for both samples with a plateau thereafter.

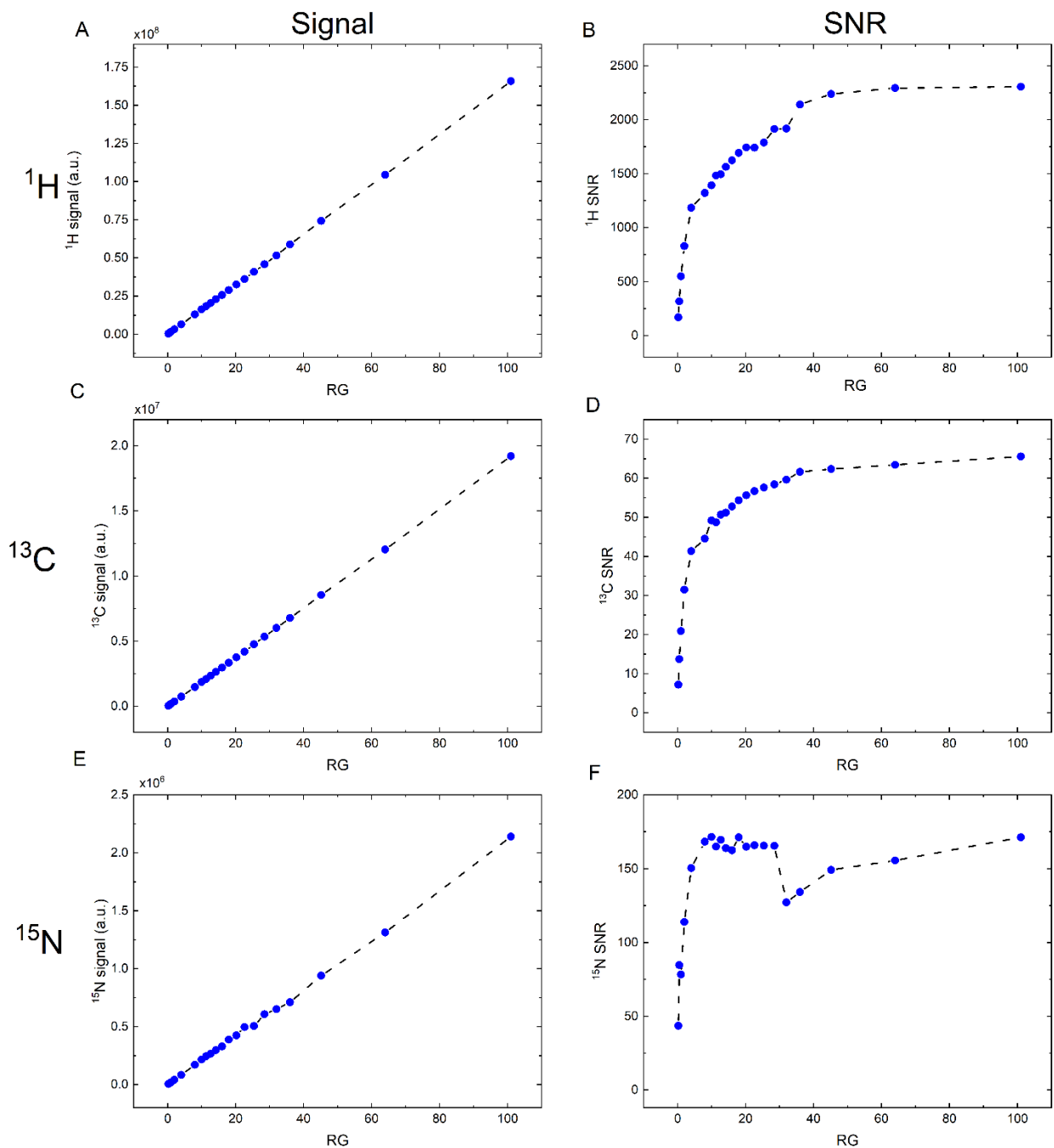

*Figure S5: Signal (left), corresponding SNR (right) of $^{13}C$ reference sample as a function of RG at 14.1 T using a Helium cryo-probe and the Avance NEO console.* The reference sample composition is given in the methods section. Note the continuous but decelerating increase of SNR from RG 0.25 to RG 101, the maximum is reached for RG 101.

# $^1$H data at 9.4 T

*Table S1:* Signal-to-noise-ratio (SNR) and maximum FID magnitude values measured with 1H sample depending on receiver gain (RG).

| RG | SNR | max FID signal (a.u.) |
|---:|---:|---:|
| 0.25 | 2.21E+03 | 2.13E+06 |
| 0.5 | 4.23E+03 | 4.24E+06 |
| 1 | 7.19E+03 | 8.51E+06 |
| 2 | 1.06E+04 | 1.70E+07 |
| 4 | 1.32E+04 | 3.40E+07 |
| 8 | 1.34E+04 | 6.79E+07 |
| 10 | 1.32E+04 | 8.54E+07 |
| 11.3 | 1.30E+04 | 9.48E+07 |
| 12.7 | 1.33E+04 | 1.07E+08 |
| 14.2 | 1.30E+04 | 1.19E+08 |
| 16 | 1.36E+04 | 1.34E+08 |
| 18 | 1.39E+04 | 1.49E+08 |
| 20.2 | 1.43E+04 | 1.69E+08 |
| 22.6 | 1.45E+04 | 1.89E+08 |
| 25.4 | 1.57E+04 | 2.13E+08 |
| 28.5 | 1.50E+04 | 2.38E+08 |
| 32 | 1.52E+04 | 2.68E+08 |
| 36 | 1.56E+04 | 3.00E+08 |
| 45.2 | 1.55E+04 | 3.78E+08 |
| 64 | 1.52E+04 | 5.34E+08 |
| 101 | 1.59E+04 | 8.47E+08 |

**Data generated from analysis:** /1H

++++++++++++++++++++++++++++++++++++++++++++++++++++++++++++++++++++

Analysis settings:

Alpha reference: 90°

Averages reference: 8

Concentration reference: 3.456 M

Polarization reference: 0.0030981%

-----

Concentration experiment: 111.02 M

Polarization experiment: 0.0030981%

Naximum ADC amplitude: 1.344E+09 a.u.

-----

Extracted RG array length: 21

Extracted SNR array length: 21

Extracted signal array length: 21

-----

Lowest RG value which yields 95% of the maximum SNR: 32

-----

Maximum SNR in the masked plot: 459230

Corresponding RG values of max SNR: 4.89379

Corresponding flipping-angle (degrees) of max SNR: 90

Maximum product of sin(alpha) and RG with chosen concentration (111.02 M) and polarization (0.0030981%) before clipping: 4.98604

-----

Your molar concentration in use was: 0.00343951

Maximum flip angle for RG 32 and molar concentration 0.00343951: 8.83768°

Maximum valid molar concentration for RG 32 and 90°: 0.00053728

++++++++++++++++++++++++++++++++++++++++++++++++++++++++++++++++++++

# $^2$H data at 9.4 T

*Table S2:* Signal-to-noise-ratio (SNR) and maximum FID magnitude values measured with 2H sample depending on receiver gain (RG):

| RG | SNR | max FID signal (a.u.) |
|---:|---|---:|
| 0.25 | 5.34E+03 | 2.32E+05 |
| 0.5 | 1.15E+04 | 4.62E+05 |
| 1 | 1.23E+04 | 9.22E+05 |
| 2 | 1.90E+04 | 1.84E+06 |
| 4 | 2.36E+04 | 3.68E+06 |
| 8 | 2.65E+04 | 7.31E+06 |
| 10 | 2.58E+04 | 9.23E+06 |
| 11.3 | 2.65E+04 | 1.04E+07 |
| 12.7 | 2.72E+04 | 1.16E+07 |
| 14.2 | 2.62E+04 | 1.30E+07 |
| 16 | 2.53E+04 | 1.46E+07 |
| 18 | 2.66E+04 | 1.64E+07 |
| 20.2 | 1.78E+04 | 1.85E+07 |
| 22.6 | 1.92E+04 | 2.06E+07 |
| 25.4 | 1.96E+04 | 2.32E+07 |
| 28.5 | 2.04E+04 | 2.61E+07 |
| 32 | 2.20E+04 | 2.93E+07 |
| 36 | 2.24E+04 | 3.28E+07 |
| 45.2 | 2.39E+04 | 4.14E+07 |
| 64 | 2.47E+04 | 5.85E+07 |
| 101 | 2.57E+04 | 9.25E+07 |

**Data generated from analysis:** /2H

++++++++++++++++++++++++++++++++++++++++++++++++++++++++++++++++++++

Analysis settings:

Alpha reference: 90°

Averages reference: 8

Concentration reference: 2.018 M

Polarization reference: 0.000475899%

-----

Concentration experiment: 110.42 M

Polarization experiment: 0.000475899%

Naximum ADC amplitude: 1.244E+09 a.u.

-----

Extracted RG array length: 21

Extracted SNR array length: 21

Extracted signal array length: 21

-----

Lowest RG value which yields 95% of the maximum SNR: 14.2

-----

Maximum SNR in the masked plot: 1.48732E+06

Corresponding RG values of max SNR: 12.768

Corresponding flipping-angle (degrees) of max SNR: 90

Maximum product of sin(alpha) and RG with chosen concentration (110.42 M) and polarization (0.000475899%) before clipping: 24.8207

-----

Your molar concentration in use was: 0.000525487

Maximum flip angle for RG 14.2 and molar concentration 0.000525487: 90°

Maximum valid molar concentration for RG 14.2 and 90°: 0.000919346

++++++++++++++++++++++++++++++++++++++++++++++++++++++++++++++++++++

# $^{13}$C data at 9.4 T

*Table S3:* Signal-to-noise-ratio (SNR) and maximum FID magnitude values measured with $^{13}$C sample depending on receiver gain (RG).

| RG | SNR | max FID signal (a.u.) |
|---:|---|---:|
| 0.25 | 4.94E+01 | 2.64E+04 |
| 0.5 | 6.07E+01 | 5.14E+04 |
| 1 | 1.10E+02 | 1.03E+05 |
| 2 | 1.70E+02 | 2.05E+05 |
| 4 | 2.15E+02 | 4.00E+05 |
| 8 | 2.25E+02 | 7.91E+05 |
| 10 | 2.27E+02 | 9.76E+05 |
| 11.3 | 2.24E+02 | 1.12E+06 |
| 12.7 | 2.27E+02 | 1.25E+06 |
| 14.2 | 2.24E+02 | 1.40E+06 |
| 16 | 2.27E+02 | 1.51E+06 |
| 18 | 2.24E+02 | 1.75E+06 |
| 20.2 | 1.53E+02 | 2.03E+06 |
| 22.6 | 1.66E+02 | 2.28E+06 |
| 25.4 | 1.79E+02 | 2.50E+06 |
| 28.5 | 1.83E+02 | 2.93E+06 |
| 32 | 1.89E+02 | 3.22E+06 |
| 36 | 1.97E+02 | 3.56E+06 |
| 45.2 | 2.06E+02 | 4.41E+06 |
| 64 | 2.13E+02 | 6.42E+06 |
| 101 | 2.21E+02 | 9.83E+06 |

**Data generated from analysis:** /13C

++++++++++++++++++++++++++++++++++++++++++++++++++++++++++++++++++++

Analysis settings:

Alpha reference: 90°

Averages reference: 2

Concentration reference: 1.308 M

Polarization reference: 0.00077917%

-----

Concentration experiment: 0.09 M

Polarization experiment: 35%

Naximum ADC amplitude: 1.244E+09 a.u.

-----

Extracted RG array length: 21

Extracted SNR array length: 21

Extracted signal array length: 21

-----

Lowest RG value which yields 95% of the maximum SNR: 8

-----

Maximum SNR in the masked plot: 60998.8

Corresponding RG values of max SNR: 9.40909

Corresponding flipping-angle (degrees) of max SNR: 5

Maximum product of sin(alpha) and RG with chosen concentration (0.09 M) and polarization (35%) before clipping: 4.1353

-----

Your molar concentration in use was: 0.0315

Maximum flip angle for RG 8 and molar concentration 0.0315: 30.5513°

Maximum valid molar concentration for RG 8 and 90°: 0.182103

++++++++++++++++++++++++++++++++++++++++++++++++++++++++++++++++++++

# $^{15}$N data at 9.4 T

***Table S4:*** *Signal-to-noise-ratio (SNR) and maximum FID magnitude values measured with $^{15}$N sample depending on receiver gain (RG).*

| RG | SNR | max FID signal (a.u.) |
|---:|---:|---:|
| 0.25 | 4.52E+01 | 4.03E+03 |
| 0.5 | 5.73E+01 | 7.29E+03 |
| 1 | 9.91E+01 | 1.42E+04 |
| 2 | 1.60E+02 | 2.69E+04 |
| 4 | 1.91E+02 | 5.32E+04 |
| 8 | 2.00E+02 | 1.02E+05 |
| 10 | 2.03E+02 | 1.31E+05 |
| 11.3 | 2.08E+02 | 1.48E+05 |
| 12.7 | 2.05E+02 | 1.65E+05 |
| 14.2 | 2.08E+02 | 1.79E+05 |
| 16 | 2.05E+02 | 2.05E+05 |
| 18 | 2.03E+02 | 2.23E+05 |
| 20.2 | 1.39E+02 | 2.63E+05 |
| 22.6 | 1.48E+02 | 2.92E+05 |
| 25.4 | 1.54E+02 | 3.35E+05 |
| 28.5 | 1.62E+02 | 3.64E+05 |
| 32 | 1.72E+02 | 4.08E+05 |
| 36 | 1.76E+02 | 4.75E+05 |
| 45.2 | 1.89E+02 | 5.71E+05 |
| 64 | 1.90E+02 | 8.32E+05 |
| 101 | 2.06E+02 | 1.26E+06 |

**Data generated from analysis:** /15N

++++++++++++++++++++++++++++++++++++++++++++++++++++++++++++++++++++

Analysis settings:

Alpha reference: 90°

Averages reference: 16

Concentration reference: 2.059 M

Polarization reference: 0.00031414%

-----

Concentration experiment: 0.04 M

Polarization experiment: 15%

Naximum ADC amplitude: 1.244E+09 a.u.

-----

Extracted RG array length: 21

Extracted SNR array length: 21

Extracted signal array length: 21

-----

Lowest RG value which yields 95% of the maximum SNR: 8

-----

Maximum SNR in the masked plot: 33507.7

Corresponding RG values of max SNR: 14.1814

Corresponding flipping-angle (degrees) of max SNR: 10

Maximum product of sin(alpha) and RG with chosen concentration (0.04 M) and polarization (15%) before clipping: 107.151

-----

Your molar concentration in use was: 0.006

Maximum flip angle for RG 8 and molar concentration 0.006: 90°

Maximum valid molar concentration for RG 8 and 90°: 0.452103

++++++++++++++++++++++++++++++++++++++++++++++++++++++++++++++++++++